\documentclass{IEEEtran}
\usepackage{cite}
\usepackage{amsmath,amssymb,amsfonts}
\usepackage{algorithmic}
\usepackage{url}
\usepackage{balance}
\usepackage{graphicx}
\usepackage{textcomp}
\def\BibTeX{{\rm B\kern-.05em{\sc i\kern-.025em b}\kern-.08em
    T\kern-.1667em\lower.7ex\hbox{E}\kern-.125emX}}
    
\renewenvironment{IEEEbiography}[1]
  {\IEEEbiographynophoto{#1}}
  {\endIEEEbiographynophoto}

\begin{document}

\title{6G Communication: Envisioning the Key Issues and Challenges}

\author{Sabuzima Nayak and Ripon Patgiri\\National Institute of Technology Silchar

\thanks{Sabuzima Nayak and Ripon Patgiri, Department of Computer Science \& Engineering, National Institute of Technology Silchar, Cachar-788010, Assam, India, Email: \textit{sabuzimanayak@gmail.com} and \textit{ripon@cse.nits.ac.in} }
}


\maketitle
\begin{abstract}
In 2030, we are going to evidence the 6G mobile communication technology, which will enable the Internet of Everything. Yet 5G has to be experienced by people worldwide and B5G has to be developed; the researchers have already started planning, visioning, and gathering requirements of the 6G. Moreover, many countries have already initiated the research on 6G. 6G promises connecting every smart device to the Internet from smartphone to intelligent vehicles. 6G will provide sophisticated and high QoS such as holographic communication, augmented reality/virtual reality and many more. Also, it will focus on Quality of Experience (QoE) to provide rich experiences from 6G technology. Notably, it is very important to vision the issues and challenges of 6G technology, otherwise, promises may not be delivered on time. The requirements of 6G poses new challenges to the research community. To achieve desired parameters of 6G, researchers are exploring various alternatives. Hence, there are diverse research challenges to envision, from devices to softwarization. Therefore, in this article, we discuss the future issues and challenges to be faced by the 6G technology. We have discussed issues and challenges from every aspect from hardware to the enabling technologies which will be utilized by 6G.
\end{abstract}

\begin{IEEEkeywords}
6G, Wireless Communication, Mobile Communication, Issues, Challenges, Internet of Things, Internet of Everything.
\end{IEEEkeywords}


\section{Introduction}
Nowadays, 6G mobile communication technology is one of the most demanded research fields. The 6G will revolutionize personal life, lifestyle, society, business and communication systems. It is a crucial time to envision the potential applications, techniques, use cases, and challenges of 6G technology. Noteworthy, 6G will create enormous research possibilities and enable many new technologies. Moreover, the 6G will be proven as a game changer paradigm in diverse fields. Therefore, the visioning of the 6G technology is necessary to revolutionize the modern world. Currently, many countries are deploying the 5G technology in full swing. However, 5G and Beyond 5G (B5G) will be unable to provide the complete requirements of the Internet of Everything (IoE). Therefore, there is a high demand for 6G. Researchers have already initiated solving the challenges of B5G mobile communication. In the next five years, B5G will evolve and it is highly expected that 6G will be fully able operate from 2030. The 6G will be the most prominent research field for next 10-25 years. The 6G research work has already begun in Finland in 2018 and the project name is ``6genesis flagship project'' \cite{Katz}. In 2019, USA, China, and South Korea have launched 6G project \cite{Dang}. Japan has also launched 6G project in 2020 \cite{Docomo}. Also, Nippon Telegraph and Telephone Public Corporation (NTT) has released their white paper in January 2020. Many countries have initiated the 6G project.

\begin{figure*}[!ht]
    \centering
    \includegraphics[width=\textwidth]{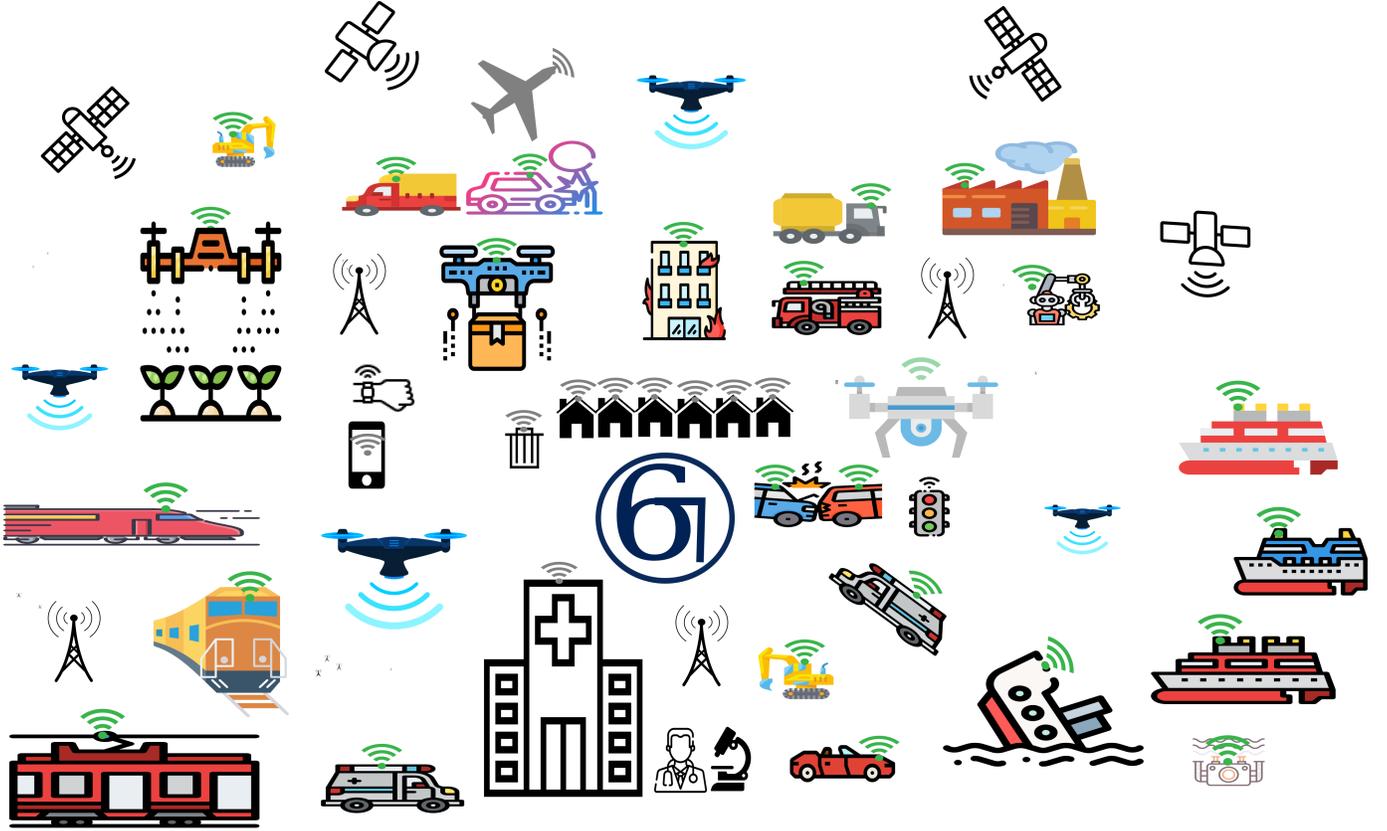}
    \caption{Envisioning the challenges of 6G}
    \label{fig1}
\end{figure*}

We are witnessing that there is a new mobile generation in every decade \cite{Chen}. Therefore, it is expected that 6G will be deployed in 2030. The maximum data rate of various mobile generation from 1G to 6G is exponential. Therefore, the 6G is itself a great challenge to achieve with all the desired parameters stated in Table \ref{tab1}. In a nutshell, the first research paper of 6G appears in 2018 \cite{Katz}. From this, research is spiking up on 6G, and gathering various requirements. Numerous discussions are performed regarding the requirements \cite{Zong,David,Zhang1}. To achieve the requirements of 6G, various visions have been provided. Gui \textit{et al.} discusses on various possible applications of mobile broad bandwidth and low latency (MBBLL), massive broad bandwidth machine type (mBBMT), massive low latency machine type (mLLMT) \cite{Gui}. Zhang \textit{et al.} focuses on further-enhanced mobile broadband (FeMBB), extremely reliable and low-latency communications (ERLLC), ultra-massive machine-type communications (umMTC), long-distance and high-mobility communications (LDHMC) and  extremely low-power communications (ELPC) \cite{Zhang1}. Dang \textit{et al.} provides future possibilities of 6Gs \cite{Dang}. Moreover, already carious works have been initiated by the researchers \cite{Al-Eryani}. Thus, there are numerous research paper in visioning the 6G mobile communication. On the contrary, it is very important to unearth the hardship to be faced by the much awaited 6G technology. Because, the 5G technology is yet to be deployed worldwide in full scale. Moreover, there are numerous issues in 5G technology to overcome. Thus, it motivated us to investigate on the issues and challenges of 6G technology.

The 6G will change the definition and perception of the modern lifestyle, society, business and communication.  It is expected that 6G will associate in revolutionizing several fields which are to be visioned. However, there are numerous problems in 6G, and thus, we exploit all issues and challenges of 6G from every aspect in this article. Moreover, we survey various requirements of 6G technology. Thus, Section \ref{req} analyzes and compares the requirements of 5G, B5G and 6G. There are many requirements to analyze and acquired before implementing the 6G. These requirements are the challenges of B5G and 6G in the implementation process. Thus, this article explores all prime challenges associated with 6G technology in Section \ref{ch}. Moreover, this article also uncovers prominent issues of 6G technology in Section \ref{iss}.  Finally, the article draws a suitable conclusion in Section \ref{con}.

\section{6G Requirements}
\label{req}

\begin{table*}[!ht]
    \centering
    \caption{Comparison among 5G, B5G and 6G requirements}
    \begin{tabular}{|p{4cm}|p{4cm}|p{4cm}|p{4cm}|}
    \hline
   \centering \textbf{Requirements} &  \centering \textbf{5G} &  \centering \textbf{B5G} &  \textbf{6G} \\ \hline
     Application types \cite{Zhang1,Saad} & $\bullet$  eMBB \newline  $\bullet$ URLLC \newline \hspace{1cm} $\bullet$ mMTC  &  $\bullet$ Reliable eMBB \newline $\bullet$ URLLC \newline $\bullet$ mMTC \newline $\bullet$ Hybrid (URLLC + eMBB)  &  $\bullet$ MBRLLC \newline $\bullet$ mURLLC \newline $\bullet$ HCS \newline $\bullet$ MPS \\ \hline
     
     Device types \cite{Zhang1,Saad} & $\bullet$ Smartphones \newline $\bullet$ Sensors \newline $\bullet$ Drones & $\bullet$ Smartphones \newline $\bullet$ Sensors \newline $\bullet$ Drones \newline $\bullet$ XR equipment & $\bullet$ Sensors and DLT devices \newline $\bullet$ CRAS \newline $\bullet$ XR and BCI equipment \newline $\bullet$ Smart implants  \\ \hline
     
     Operating frequency \cite{Zhang1}  & 3-300 MHz & 500 MHz & 1 THz \\ \hline
     
     Spectral efficiency \cite{Zhang1,Saad} & 10$\times$ in bps/Hz/m$^2$/Joules & 100$\times$ in bps/Hz/m2/Joules & 1000$\times$ in bps/Hz/m$^3$/Joules\\ \hline
     
     Data Rate \cite{Zhang1,Saad} & 1 Gbps & 100 Gbps & 1 Tbps \\ \hline
     
     End-to-end delay  \cite{Zhang1,Saad} & 5 ms & 1 ms & $<$1 ms\\ \hline
     
     Radio-only delay  \cite{Zhang1,Saad} & 100 ns & $<$100 ns & $\leq$10 ns\\ \hline
     
     Processing delay \cite{Zhang1,Saad} & 100 ns & 50 ns & $\leq$ 10 ns\\ \hline
     
     Mobility range \cite{Zhang1} & 100 -- $<$ 500 km/h & 500 km/h& 1000 km/h \\ \hline
     
     Wavelength & 3 mm & 1 mm & 300 $\mu$m\\ \hline
     
     Architecture & Massive MIMO & Massive MIMO & Intelligent surface \\ \hline
     
     Core network & Internet of Things & Internet of Things & Internet of Everything \\ \hline
     
     Satellite integration & Partial & Partial & Full \\ \hline
     
     Artificial Intelligence integration & Partial & Full & Truly Artificial Intelligence Driven \\ \hline
     
     XR integration & Partial & Full & Full \\ \hline
     
     Haptic communication integration & Partial & Full & Full \\ \hline
     
     Highlights & Extremely high rate of Streaming & Extremely high rate of Streaming & Security, secrecy, \& privacy \\ \hline
     
    \end{tabular}
    \label{tab1}
\end{table*}

Table \ref{tab1} depicts the requirements and the challenges of 6G. The 6G is expected to deliver truly Artificial Intelligence (AI)-driven communications. Therefore, ``the 6G is about the 6th sense”, Peter Wetter, Nokia Bell Labs \cite{Viswanathan}. Also, it is expected that prices will be $1000\times$ cheaper in 6G era \cite{Zhang2}. Zhang \textit{et al.} \cite{Zhang2} emphasizes price reduction that there were $10\times$ price reduction from 1G to 2G, and $1000\times$ in 3G and 4G. It will be expected to continue in 5G and 6G era. Moreover, 6G will be backed by Extended Radio (XR) and will replace NR-Lite of 5G. Also, it requires satellite support with terrestrial communication. The prime requirements of 6G technology are 1 THz operating frequency and data rate of 1 Tbps. Also, the 6G is 3D structural communication which defines space, time and frequency with requirements of $100\times~in~bps/Hz/m^3/Joules$. It expects maximum wavelength of $300~\mu m$ to achieve 1 THz. 

\subsection{Network Architecture}
Most importantly, 6G technology will enable the IoE. Every smart device will be connected in the 6G network. Starting from small devices such as smartwatch or smart phones to mobile healthcare, vehicular and so on. IoE and mMTC services will connect all smart devices to one or more wireless-access networks. These wireless-access networks have multiple access points (AP)/base stations (BS) to provide services. To provide services to million smart devices within a small geographical location, the APs/BSs are located densely/ultradensely in 6G network. In such a network scenario, APs/BSs will have overlapping coverage areas, i.e., a small geographical location will be served by multiple APs/BSs. And, multipoint transmissions will have million users served by multiple APs/BSs. In this case, management and efficiency are required in frequency allocation, interference management and efficient handoff. APs/BSs are connected through fast backhauling links. The 6G network will be viewed as a distributed, cell-less massive multiple-input, multiple-output (MIMO) system. In this network, every smart device will be served by more than one APs/BSs using transmission multiplexing or transmission coordination. The APs/BSs are connected to the Cloud for accessing Cloud services. Cloud computing supports large smart applications such as autonomous driving, virtual reality, smart city, industrial control and smart manufacturing, and numerous health-monitoring services. The smart devices are connected using peer-to-peer connectivity through single/multihop communications. Moreover, terrestrial cellular systems will be integrated with drone-assisted/aerial networks supported by mobile APs. In addition, these networks will be based on content and application driven more than just data transmission.

\section{Challenges}
\label{ch}
The challenges are much bigger than 6G services deliverable as depicted in Figure \ref{fig1} through an iceberg. It is expected that most of the issues of 5G will be solved by B5G and the remaining issues by 6G. However, the requirements are growing exponentially. Thus, challenges of 6G are much bigger than the 5G and B5G.

\subsection{Terahertz Signal}
6G will be using TeraHertz (THz) frequency band for transmission. However, operating in THz is a grand challenge where the wavelength is 300 $\mu$m. 
\begin{itemize}
    \item \textbf{Generation}: 6G transmission requires continuous THz signal. But, generation of continuous THz signal is difficult because it has more strict requirements regarding size and has more complexity in designing the antenna/transmitter. Moreover, it is costly to generate of THz signal. Transmission aspect of 6G has to be low in cost to support the low cost advantage claimed by 6G. 
    
    \item \textbf{Energy Loss}: Another issue with THz signal is that it is a short distance communication signal (i.e., a few meters). Because, it attenuates to zero after travelling a short distance in air. The energy loss of signal occurs due to molecular absorption and spreading loss. The molecular absorption loss occurs due to conversion from THz signal energy to the internal kinetic energy of the molecules present in the air. And, the loss becomes severe with increasing moisture in the air. The spreading loss occurs due to extension of electromagnetic waves in the air and quadratically increases based on the operating frequency and distance between two communication ends. This is a major setback for 6G because such a signal is inappropriate for 6G requirements. It is not economical to amplify a signal after every 1 meter. However, research continues to make THz signal ideal for 6G. 
   
\end{itemize}

\subsection{New Communication Alternatives}
New communication alternatives are also explored to have a better option instead of THz signals. Visible Light Communication (VLC) uses cheap light emitting diodes (LED) to achieve higher frequency bands \cite{Huang}. However, the VLC has issues with coverage and noise interference from other sources of light. Thus, VLC is used in a confined arena that does not have any interference from other sources of light. Another option is molecular communications (MC) \cite{Huang}. It uses biochemical signals for data transmission. Biochemical signals are particles of the size of a few nanometers to a few micrometers. It propagates in a gaseous or aqueous medium. Advantages of MC signals are biocompatible, consumes less energy for production and transmission, and high data rates. However, it creates challenges of security and effective interface between chemical and electrical domains. Third option can be quantum communication (QC) \cite{Huang}. Quantum particles or photons are utilized to encode data in a quantum state. It makes the data access and cloning by hackers difficult. The advantages are high security, high data rates, and effective long distance transmission. However, its at an early stage of development and have a long way to be considered as an alternative of THz signal. 

\subsection{Underwater Communication}
6G is also aiming to provide underwater communication. However, the underwater environment is becoming a different scenario compared to air or space. Underwater environment is unpredictable and complex, such as high signal attenuation, physical damage to equipment and complicated network deployment. Radio signals are highly attenuated in salt water. Therefore, acoustic communication is the only option for communication. Deployment of nodes is expensive and difficult, hence, density of nodes have to be low. Node mobility is difficult due to flow and density variation of the water. Underwater sensors are expensive and these sensors are designed with extra protection for the underwater environments. Moreover, it requires complex transceivers and a large memory. The power supply has to be large because solar power cannot be utilized. Fouling and corrosion increase the failure of underwater sensors. The best choice is optical fiber, however, it is expensive. Another option is quantum communication, however, it is at an early stage of research. To achieve an efficient underwater communication 6G has to fight against the challenges of underwater circumstances.

\subsection{Design of Transmitter and Antenna}
The 6G demands highly efficient transmitter and receiver antenna. The required radio frequency (RF) transceiver should have high integration. It is achieved by implementing advanced silicon nodes with CMOS SOI, bulk CMOS and SiGe BiCMOS and attentive co-design with off-chip, highly efficient antennas. To meet the requirements of 6G, new balance between RF technologies, communication and signal processing is essential \cite{Katz}. Highly efficient RF will help in achieving high data rates. In addition, the materials used to construct the antennas also influence the data rate. The materials influence the intrinsic and extrinsic composition-structure-property relationships. If this relationship is poor, then the attenuation is more and leading to low data rate. Currently, research on new materials are being conducted such as nanomaterials, bio-based, foams, room-temperature fabricated materials, ultra-low permittivity. These materials are self healing and electrically tuneable \cite{Katz}. 

In case of satellite assisted IoE, the IoT devices pose the most issues. In satellite assisted IoE, first, data are collected from IoT devices which are transmitted to satellite antennas at the ground connected to core networks through wireless or wired channels. However, IoT devices have hardware constraints. Therefore, data transmission to satellites is difficult. Suppose, 6G follows a direct transmission mode. It means that the IoT devices directly send the data to satellites. In such a case, IoT devices have to provide high power to the signal to reach the satellite, however, it is impractical due to low power source of IoT devices. Another solution is hybrid transmission mode. A data aggregator collects the IoT data and transmit to the satellite. 6G requires mitigation/interference coordination techniques along with varieties of MIMO multi-antenna transceivers for good transmission \cite{Katz}. 

\subsection{End-to-end Delay and Reliability}
The 6G data rate requirement is 1 Tbps which will enable many applications, to name a few, virtual/augmented reality, smart healthcare, UAV, smart electric car, and smart city. Such applications require high data rate with low latency. Notably, 6G focuses on QoE. For instance, a car discusses with passengers, which enable the teaching - learning process. Also, electric car need to charge while moving without the intervention of the human. To provide a high QoE, it demands low end-to-end delay. To achieve lower end-to-end delay, smaller frame sizes or data packets have to be transmitted. Moreover, 6G will require flat network architectures. Reliable transmission requires efficient Forward error correction (FEC) schemes. Efficient FEC schemes requisite long transmission time or using parallel diversity channels in large numbers \cite{Katz}. Long transmission time is not possible because 6G has to maintain a latency of $<$ 1 ms. This challenge need to be addressed in physical and networking layers. Technology such as AI is also considered to reduce latency, for example, deep learning based transmission prediction. It helps in predicting the user requests and change in channel state to reduce the transmission latency. The 5G is backed by ultra-reliable low latency communications (URLLC) and it's reliability is $99.999\%$. On the contrary, 6G has to increase its reliability and requires extremely reliable low latency communications (ERLLC) to support a reliability rate of $99.99999999\%$. Therefore, 6G has to enhance URLLC and provide higher reliability rate than 5G and B5G.

\subsection{Energy}
Every device in 6G, such as smart devices and APs, will implement sophisticated signal processing mechanisms. Moreover, they have to process Big data which require heavy processing and high energy. Therefore, energy is also an issue in 6G. 6G will be using emerging technology such as Edge and AI in its network nodes that also require high energy. Thus, 6G has to solve the issue of harvesting, charging and conservation of energy. Another factor is energy cooperation among 6G network nodes. In addition, data transmission also demands more energy as discussed in the transmission mode. Therefore, to reduce the power consumption, new waveform and modulation is required to have low peak to average power ratios (PAPR) \cite{Chen}. Besides, higher energy is consumed to compute complex algorithms to provide high level security. On the contrary, embedding AI with 6G will reduce the energy consumption.

\subsection{Capacity}
6G is a key enabler of IoE that connects billions of smart devices and smart wearables. Moreover, 6G promises many high QoS applications, e.g, virtual/augmented reality. The IoE devices will produce huge traffic. But, the aim of $<$1 ms latency is achievable in case of low traffic. Moreover, communication network congestion will degrade QoS. Therefore, capacity is a big challenge for 6G. 6G capacity is enhanced mainly by four ways. First, increase the spectrum bandwidth. Second, enhance the spectrum efficiency in the air, i.e., the bits per second per Hertz. It is achieved by implementing good channel coding or new modulation technique. Third, spectrum reuse. Another solution is increasing the node density. Mobile nodes are deployed to hotspot network areas to reduce the load of the nodes. Nevertheless, the repositioning of mobile node affects many factors such as handover control and network topology. Manually handling the repositioning is difficult, hence, AI is deployed.   

\subsection{Global Coverage}
6G will rely on the low-earth orbit (LEO) satellite having a height of 500 to 2000 km from the orbit for providing global coverage. LEO will aim for less path loss, lower transmission delay and much more. However, LEO have issues such as doppler variation, doppler shift, long transmission delay and more path loss \cite{Chen}. LEO satellites travel very fast compared to the rotation of Earth. This leads to doppler variation and doppler shift in the network communication and cause random access, synchronization, signal detection, and signal measurement issues. Other issues are long transmission delay and more path loss in LEO compared to terrestrial transmission. 

\subsection{Density}
Density is the number of network nodes per square kilometer. 6G will have more density to have a smooth and high QoS. However, deploying a number of network nodes will increase the communication cost. Communication cost refers to the congestion control, scheduling, synchronization, failure detection and many more. The economic cost will also be high because 6G include both non-terrestrial and terrestrial nodes. The non-terrestrial nodes are very costly compared to terrestrial nodes. Moreover, more nodes require more maintenance.

\subsection{Cost}
One of the goals of 6G is providing the services at a cheaper cost. The 6G network nodes are both non-terrestrial and terrestrial. Terrestrial nodes are comparatively low in cost, however, their maintenance will incur money. The non-terrestrial nodes, such as satellites, drones and other mobile nodes, are very expensive. Deploying satellites to space is a costly affair. Their repair and maintenance is another issue in terms of cost. In addition, underwater communication also requires expensive infrastructure. Moreover, the 6G requires high quality devices to maintain the highest QoS. Sophisticated and high quality devices are costly. The smart devices are costly and may not be affordable to everyone. The challenge is to reduce to cost to make affordable to everyone.

\subsection{Heterogeneity}
The 6G will connect a variety of smart devices. Moreover, the communication network will be different at different coverage. It is impossible to provide just a single global coverage to connect the whole world. The communication network will be divided into sub-networks. And, these sub-networks will not be homogeneous. In addition, 6G will integrate both non-terrestrial and terrestrial communication networks, whereas both are very heterogeneous by nature. Their different heterogeneous features have to be considered to efficiently integrate them together. Heterogeneity is also present in the protocol that these communication networks will follow.  Thus, 6G is taking a big challenge of integrating various heterogeneous components together. 

\subsection{Security, Secrecy and Privacy}
Dang \textit{et. al} \cite{Dang} propose that the security, secrecy and privacy are the key features of 6G. It is expected that 6G will be able to deliver the highest level of security. THz communication will make the 6G eavesdropping and jamming proof. Similarly, quantum computing will feature the unbreakable security to the 6G, for instance, quantum cryptography for secrecy. The privacy is the key concerns of individuals, particularly, healthcare requires the highest degree of privacy protection. Blockchain is the most prominent technology to achieve a higher degree of privacy, secrecy and security. Deploying federated AI also increases security. Another important point to note is 6G promises to provide physical layer security. It will be achieved by integrating AI to 6G. Various research works are going on to explore this option \cite{Al-Eryani}. Another vulnerable target of 6G is the power system of the network devices. For DoS attacks, attackers are adopting sleep deprivation or battery draining. To achieve sleep deprivation attacks the attacker transmits many requests to Edge devices or other network nodes pretending as a legitimate user. The nodes will respond to the requests and exhausts the power. In battery draining attack, using any means the attacker tries to exhaust the power. For example, executing power consuming subroutines in the 6G network nodes. Thus, such simple trick will be able to cut off the network nodes and will impair the communication. 

\subsection{Heavy Computation}
6G will combine communication with computation. The services of 6G are heavy computation. 6G networks are ultra large scale, complex and multidimensional. It is also dynamic due to network topology, consumer demands, traffic load and radio resource. Thus, automatic network configuration is required for wireless connection. Similarly, the 6G network requires intelligent mobility management. It helps in positioning the mobile node during a disaster or node failure. However, all these tasks require heavy computation. To solve that 6G will rely on new technologies such as edge computing, federated AI, etc. However, implementing these technologies also has many sets of issues. Issues are present both in individual technology and integration with 6G. 

\subsection{Five Senses Communication}
The 6G is a sixth sense communication system. Therefore, it is expected that 6G will be able to support five senses of communication system. The five senses are sight, hearing, smell, touch and taste. 6G requires a very high data rate with extremely low latency. Moreover, the sensors must able to reproduce the five senses from remote locations to provide real experience to the users, for example, holographic communication. Therefore, 6G is focused on QoE.

\subsection{Industrial Revolution}
We expect that Industry 5.0 will evolve and revolutionized soon. There is a high chance of shifting of Industry 4.0 to Industry 5.0 in 2030. Industry 4.0 is ``\textit{Digitalization}'', and Industry 5.0 is ``\textit{Personalization and Intelligentization}'' which requires 5G, B5G and 6G. Industry 5.0 will allow the customer to customize the services. 6G will enable Industrial Internet of Everything (IIoE), and further, it will revolutionize the industries a step ahead and motivate to introduce ``\textit{Intelligentization}''. Moreover, we foresee that 6G and B6G will be a great enabler of Industry 5.0 which is ``\textit{Intelligentization}'' of industries. 


\subsection{Paradigm Shift}
6G will enable paradigm shift and we will evidence the shift from 2030 onward. Furthermore, 6G will enable the conversion from smart devices to intelligent devices, for instance, intelligent vehicle, intelligent mobile, intelligent wearables, intelligent healthcare and many more. It will be possible with the combination of AI and 6G communication technology. However, integration of intelligence and the conversion from smart to intelligent devices involve many issues. Therefore, new sensors, new devices and new methods have to be devised to achieve this challenge.

\section{Issues}
\label{iss}
Apparently, the 5G technology has just begun to deploy. Also, there is no user experiences on B5G and the B5G has to experience many things in real scenario than a lab. Moreover, B5G mobile communication is yet to be developed completely. Therefore, there are many issues while envisioning 6G technology.

\subsection{Lack of Technology}
Fitzek and Peeling \cite{Fitzek} argues that there is a lack of real technology to implement the 6G requirements. The 6G has many promises to deliver, but hindered by underdeveloped real technology which could drastically advance 6G technology from 5G. The requirements increase exponentially from 5G to 6G mobile communications. 6G has no underpinning technology, and therefore, it poses doubt on the envisioning in this stage \cite{Fitzek}. The 5G technology has just begun. Thus, deficient in wider user experiences of 5G poses many issues. Obviously, networking can be effectively utilized using softwarization. Huge modifications are required on Software-Defined Radio, Software-Defined Network and Network Function Virtualization of 5G technology to achieve 6G \cite{Fitzek}. Also, the AI has to be embedded in 5G technology to advance 6G. The 6G is truly AI-Driven communication technology. Therefore, the AI has to be deployed in every layer of 5G. Noteworthy, 5G will partially support AI to communicate. 

\subsection{Over Expectation from 5G}
Another key issue is the exponential difference of 6G with its predecessor. The 5G is an enhanced version of its predecessor and not exponential growth of requirements. However, 4G is modified immensely to introduce 5G as uninterrupted mobile communication. We acknowledge Fitzek \textit{et al.}, the issues of 5G have to be solved before developing 6G technology. Currently, the 5G is deployed as a campus solution in a local area \cite{Fitzek}. Moreover, the 5G does not support high mobility. As a consequence, the 6G mobile communication technology will face a huge problem in providing the mobility support 1000 km/h. Also, this issue can delay the deployment of 6G to 2032 or beyond. Moreover, current state-of-the-art technology is unable to provide network coverage to the passengers in an aircraft due to various issues \cite{Dang}. But, high mobility is the key issue. This issue can be addressed by satellite communication, however, it is very costly. Therefore, another alternative is required to solve the issue. It is expected that 6G will solve the issue. However, there is no clear roadmap to provide low cost connectivity to the passengers with 6G parameters.

\subsection{Interoperability of protocols}
6G network will integrate both non-terrestrial and terrestrial communication networks. The TCP/IP protocols used in terrestrial communication networks are unsuitable for non-terrestrial communication networks, i.e., satellite communication. Because, efficiency of TCP/IP protocols is lower due to long transmission delay, ultra wide bandwidth and higher bit error rate. Therefore, modification in TCP/IP protocol is required to support both non-terrestrial and terrestrial communication networks with efficiency. Moreover, different protocols are implemented in both types of communication networks. In that case, interoperability of protocols will be an issue \cite{Chen}.  

\subsection{Artificial Intelligence}
6G will be truly AI-driven mobile wireless communication, i.e., communication system will be intelligent enough to route the data. Moreover, federated AI will help in knowledge sharing among the intelligent devices. Furthermore, quantum machine learning will enhance the performance of 6G by intelligent data analysis. However, most of the techniques are in the initial phase of research. All AI algorithms have high computation. The high computation task has a long execution time and consumes more power. Whereas, 6G is unable to provide such relaxation. Another issue is the dynamic nature of the network. Once inferences are obtained by AI algorithms, it is used to predict future incoming data. However, communication networks are very dynamic, therefore, the inferences will become obsolete quickly. Training the AI module after a short interval will be very costly for 6G. Therefore, 6G will face huge challenge to deploy AI in the communication. Furthermore, physical layer will be assisted by AI. However, implementing AI in physical layer is difficult due to the complexity of the physical layer and bounded learning capacity of AI algorithms.


\subsection{Big Data}
The 6G will enable IoE and produce massive amounts of small sized data. These Big data are stored, process and managed through Cloud Computing. The Data on Internet of Everything (DIoE) will pose new challenges to handle these data. Till date, there is no such technology that can store, process and manage Exabytes of small data. These small data will form data silo, and thus, Big Data is to be redefined and Big Data 2.0 will be introduced. However, Big data 2.0 require a supercomputer to compute and analyze data \cite{Nayak1}. 

\subsection{Edge Technology}
Edge technology was introduced because Cloud is far from the devices that generate data. However, Edge technology is also an emerging research area containing many issues. Edge nodes have small power sources and memory. Edge analytics mostly rely on AI. However, due to the incapability to execute heavy computation, the AI algorithm is executed in the Cloud and the inferences are transmitted to Edge nodes. But, the dependence of the Edge on the Cloud will reduce the performance of 6G. Moreover, the issues of AI will influence the performance of Edge, and then it will adversely influence the performance of 6G. As discussed in the Big Data section, the massive data will be stored in Edge nodes. 6G is planning to have a high density. However, with limited storage, Edge nodes will be incapable of storing information and have to transmit the data to the Cloud within a very short time interval. Thus, communication cost will be incurred upon data transmission to the Cloud. 

\section{Conclusion} 
\label{con}
6G is an important communication technology that will enable many new technologies, for instance, holographic communication. Therefore, we have surveyed the desired parameters of 6G technology. Also, we have compared the 5G, B5G and 6G mobile communication technology to depict the differences and needs of the 6G technology. However, the differences are significant. Therefore, there are huge challenges of 6G to achieve the desired parameters and deliver the promises of 6G. All issues and challenges of 6G have been uncovered from every aspect. Finally, we conclude that 6G will revolutionize many areas, and will also be proven as game changing technology in diverse fields.

\balance

\bibliographystyle{IEEEtran}
\bibliography{mybib}
\begin{IEEEbiography}{Sabuzima Nayak}
is pursuing PhD at the department of Computer Science \& Engineering, National Institute of Technology Silchar. She has published several papers in conferences and journals. She is authoring a Book. Her research interests are Bloom Filter, Communication, Big Data and Bioinformatics. 
\end{IEEEbiography}
\begin{IEEEbiography}{Ripon Patgiri} is currently working as an Assistant Professor, National Institute of Technology Silchar. He has received his PhD degree from National Institute of Technology Silchar and M.Tech. Degree from Indian Institute of Technology Guwhati. He has published several papers in reputed journals, conferences and books. His research Interests are Bloom Filter, Communication and Networking, Big Data and Bioinformatics. He is a senior member of IEEE. He has chaired and organized several international conferences. URL: http://cs.nits.ac.in/rp/
\end{IEEEbiography}
\end{document}